\documentclass[12pt]{article}
\usepackage{amssymb,amsmath}
\begin{document}

\sloppy
\title
{\Large How much weigh a pound in gravitational field   }

\author
 {
       A.I.Nikishov
          \thanks
             {E-mail: nikishov@lpi.ru}
  \\
               {\small \phantom{uuu}}
  \\
           {\it {\small} I.E.Tamm Department of Theoretical Physics,}
  \\
               {\it {\small} P.N.Lebedev Physical Institute, Moscow, Russia}
  \\
 }
%
\maketitle
\begin{abstract}
The force in the equation of motion of a particle should be in accordance with energy conservation in a constant gravitational field. It turns out that this is possible only if the force is given by the change of momentum per unit of coordinate (not proper) time. We discuss the consequences of this fact. In particular it turns out that the spring balance, which keeps the body at rest in (strong) gravitation field, shows the finite value even at infinitely close approach to horizons of black holes, if they exist in Nature.
\end{abstract}
There are several ways of introducing force in the equation of motion of a particle in gravitational field.
So M\"oller elaborates three forces "to describe all aspects of nongravitational interaction", see \S 10.4 in [1]. Landau and Lifshitz define the gravitational force as 3-space covariant derivative of momentum over (synchronized) proper time, see \S   88, Problem 1 in [2]. Not all of these forces respect the energy conservation law. Those which do it are not covariant. We consider some of these approaches.

In a constant gravitational field the covariant component of momentum $p_0$ is conserved:
(see \S  88 in [2])
$$
p_0=\frac{mc^2\sqrt{g_{00}}}{\sqrt{1-\frac{v^2}{c^2}}},\quad
v^{\alpha}=  \frac{cdx^{\alpha}}{\sqrt{g_{00}}(dx^{\alpha}-g_{\alpha}dx^{\alpha})},\quad
g_{\alpha}=-\frac{g_{0\alpha}}{g_{00}},\quad v^2=\gamma_{\alpha\beta}v^{\alpha}v^{\beta}.  \eqno(1)
$$
For particle at rest
$$
p_0=mc^2\sqrt{g_{00}},\quad dp_0=\frac{mc^2}{2\sqrt{g_{00}}}g_{00,\alpha}dx^{\alpha}.
                                                                           \eqno(2)
$$
Hence the coordinate nongravitational force, which adiabatically changes $p_0$ is
$$
\stackrel{n.g.}F_{\alpha}=\frac{dp_0}{dx^{\alpha}}=\frac{mc^2}{2\sqrt{g_{00}}}g_{00,\alpha}. \eqno(3)
$$
When only space coordinates are transformed,  $F_{\alpha}$ is 3-vector and $dp_0$ is an invariant and as such it can be integrated, see equation (10.116) in [1].

Next we assume that in a small region of space the gravitational force is directed against  the axis 1.
Then only $dx^1$ is nonzero:
$$
 dp_0=\frac{mc^2}{2\sqrt{g_{00}}}g_{00,1}dx^{1}=\frac{mc^2}{2}\frac{g_{00,1}}{\sqrt{|g_{00}g_{11}|}}dl^1, \quad dl^1=\sqrt{g_{11}}dx^1.                                           \eqno(4)
$$
 From here  it is natural to assume that  the 3-invariant nongravitational force, holding particle at rest in constant gravitational field, is the force measured by a spring balance:
$$
F=\frac{mc^2g_{00,1}}{2\sqrt{|g_{00}g_{11}|}}.                                                              \eqno(5)
$$
According to heuristic approach to gravity [3] one may expect that $|g_{00}g_{11}|=1$.
In general relativity this is true only in linear approximation in isotropic coordinates
in Schwarzschild solution, which has the form
 $$
 ds^2=g_{00}(dx^0)^2+g_{11}[(dx^1)^2+(dx^2)^2+(dx^3)^2].            \eqno(6)
 $$
 In any case $|g_{00}g_{11}|$ is always finite. Then it follows
from (6) that the invariant force acting on particle at rest remains finite even when the particle
is approaching the black hole horizon. This remains true also when $g_{0\alpha}$ are nonzero because the Coriolis-type force is not acting on particle at rest.

On the other hand according to eq. (2.2.6) in [4] when $v^{\alpha}=0$ the  gravitational force (and nongravitational one holding the particle at rest) is
$$
K= m\sqrt{\gamma_{\alpha\beta}a^{\alpha}a^{\beta}},\quad a^{\alpha}
=\frac{d^2x^{\alpha}}{ds^2},\quad \alpha=1,2.3.                                                                   \eqno(7)
$$
For the metric (6) we find
$$
K=\frac{mc^2}{2}\frac{g_{00,\alpha}}{\sqrt{|g_{00}g_{11}|}}\frac{1}{\sqrt{g_{00}}}                           \eqno(8)
$$
Contrary to (5) this quantity grows unlimitedly at the approach to black hole horizon. The contradiction
with (5) disappears if the derivative of momentum over the proper time is replaced by the derivative over
coordinate time.

The derivative over coordinate time naturally appears in Lagrange formalism, see \S 10.4 in [1]. Indeed, using
$$
L=-mc(g_{ik}\frac{dx^i}{dt}\frac{dx^k}{dt})^{1/2},                                                             \eqno(9)
$$
 we have
 $$
 p_{\mu}=\frac{\partial L}{\partial\tilde  u^{\mu}}=-\frac{mcg_{\mu l}\tilde u^l}{\sqrt{g_{ik}\tilde u^i\tilde u^k}},\quad \tilde u^k=\frac{dx^k}{dt},\quad \frac{\partial L}{\partial x^{\mu}}=-\frac{mcg_{lm,\mu}\tilde u^l\tilde u^m}{2\sqrt{g_{ik}\tilde u^i\tilde u^k}}.
                                                                                                                   \eqno(10)
 $$
 Then the Euler- Lagrange equations are
 $$
 \frac{dp_{\mu}}{dt}=\frac{\partial L}{\partial x^{\mu}}=-\frac{mcg_{lm,\mu}\tilde u^l\tilde u^m}{2\sqrt{g_{ik}\tilde u^i\tilde u^k}}=
 \stackrel{g.}F_{\mu}.                                                                                               \eqno(11)
 $$
For particle at rest $\tilde u^i=c\delta^i_0$ and $\stackrel{g.}F$ agrees with (3).( for slowly moving particle $\stackrel{g.}F_{\mu} \approx-\stackrel{n.g.}F_{\mu}$).   
Introducing a nongravitational force we have
$$
\frac{dp_{\mu}}{dt}=
 \stackrel{g.}F_{\mu} +\stackrel{n.g.}F_{\mu}.                                                                           \eqno(12)
 $$
. 

 Now we dwell briefly on Problem 1 in \S 88 in [2]. In the second edition of this book (Moscow, 1948) the problem is treated for small values of velocity. The force is obtained in the form
 $$
 f_{\alpha}=\frac{dp_{\alpha}}{d\tau}=mc^2[-\frac12\frac{g_{00,\alpha}}{g_{00}}+\sqrt{g_{00}}(g_{\beta,\alpha}-
 g_{\alpha,\beta})\frac{v^{\beta}}{c}], \quad g_{\alpha}=-\frac{g_{0\alpha}}{g_{00}}. \eqno(13)
 $$
 Replacing  the differentiation over $\tau$ by differentiation over $t$ i.e. multiplying both sides of (13) by $\sqrt{g_{00}}$
we get
$$
\sqrt{g_{00}}f_{\alpha}=\stackrel{g.}F_{\alpha}=mc^2[-\frac12\frac{g_{00,\alpha}}{\sqrt g_{00}}+g_{00}(g_{\beta,\alpha}-
 g_{\alpha,\beta})\frac{v^{\beta}}{c}]. \eqno(14)
 $$
 For particle at rest  this agrees with  (3). For arbitrary velocity the equation (3) in Problem 1 in \S 88 in [2] should agree with
 equations (10.75)-(10.77) in [1] up to a common factor, but this is not seen.

It is worthwhile to note that by itself the equivalence principal and the requirement of general covariance
do not give the equation (3) or  (5), which gives the weight of a body in constant gravitational field. So it is not a sin to try to find an algorithm for the metric, which is different from Einstein equations.

In conclusion I thank V.I.Ritus and M.I.Zelnikov for stimulating discussions.

  \section{References}
  1.C. M\"oller, {\sl The Theory of Relativity}, Clarendon Press, Oxford (1972).\\
  2.L.D.Landau and E.M.Lifshitz, {\sl The classical theory of
 fields}, Moscow, (1973) (in Russian).\\
  3. H.Dehnen, H.H\"onl, and K.Westpfahl, Ann.  der
Phys. {\bf6}, 7 Folge, Band 6, Heft 7-8, S.670 (1960).  \\
 4.I.D.Novikov, V.P.Frolov,
  {\sl Physics of black holes}, Moscow (1986) (in Russian).\\

\end{document}